# Mechanical Properties of Phosphorene Nanotubes: A Density Functional Tight-Binding Study

V. Sorkin[*] and Y.W. Zhang[†]

Institute of High Performance Computing, A*STAR, Singapore 138632

## Abstract

Using density functional tight-binding method, we studied the elastic properties, deformation and failure of armchair (AC) and zigzag (ZZ) phosphorene nanotubes (PNTs) under uniaxial tensile strain. We found that the deformation and failure of PNTs are very anisotropic. For ZZ PNTs, three deformation phases are recognized: The primary linear elastic phase, which is associated with the interactions between the neighboring puckers, succeeded by the bond rotation phase, where the puckered configuration of phosphorene is smoothed via bond rotation, and lastly the bond elongation phase, where the P-P bonds are directly stretched up to the maximally allowed limit and the failure is initiated by the rupture of the most stretched bonds. For AC PNTs, the applied strain stretches the bonds up to the maximally allowed limit, causing their ultimate failure. For both AC and ZZ PNTs, their failure strain and failure stress are sensitive while the Young's modulus, flexural rigidity, radial Poisson's ratio and thickness Poisson's ratio are relatively insensitive to the tube diameter. More specifically, for AC PNTs, the failure strain decreases from 0.40 to 0.25 and the failure stress increases from 13 GPa to 21 GPa when the tube diameter increases from 13.3 Å to 32.8 Å; while for ZZ PNTs, the failure strain decreases from 0.66 to 0.55 and the failure stress increases from 4 GPa to 9 GPa when the tube diameter increases from 13.2 Å to 31.1 Å. The Young's modulus, flexural rigidity, radial and thickness Poisson ratios are 114.2 GPa, 0.019 eV·nm$^2$, 0.47 and 0.11 for AC PNTs, and 49.2 GPa, 0.071 eV·nm$^2$, 0.07 and 0.21 for ZZ PNTs, respectively. The present findings provide valuable references for the designs and applications of PNTs as device elements.

## 1. Introduction

[*] Email address: sorkinv@ihpc.a-star.edu.sg
[†] Email address: zhangyw@ihpc.a-star.edu.sg

Phosphorene, a two-dimensional (2D) form of black phosphorus, has attracted particular interest lately due to its direct band-gap semiconducting features [1–3]. At first, multi-layer phosphorene was attained through mechanical exfoliation [1,4,5], and then monolayer phosphorene was obtained by plasma thinning process [6]. Phosphorene is a very appealing material in addition to graphene [7,8] boron nitride [9–11], and transition metal dichalcogenides (TMDs) [12–15]. Phosphorene is ideally suitable for field-effect transistors [16–21], since it a direct band-gap semiconductor with a considerable fundamental band gap. Besides its direct band gap, phosphorene has a comparatively high direction-dependent carrier mobility, which can be significantly modified by applied straining [19,22–31]. Its thermal transport [32,33] and mechanical properties are also highly anisotropic due to its puckered structure [34–36]. The wide diversity of electronic, mechanical and thermal properties under strain [18,30,34,37,38] make phosphorene a particularly appealing 2D material for strain engineering and flexible electronics [22,34,39,40]. In addition, other fascinating applications of phosphorene have been demonstrated, such as gas sensors [33,41–43], thermo-electrics [44–47], anodes in Li-ion batteries [45,48,49], p-n junctions [50–52], photo-catalysts [24] and components of solar-cell devices [20,23,53].

Along with phosphorene, of prime importance are phosphorene-based nanostructures, which are potential building blocks for a large variety of technological applications from sensing devices to active optoelectronic elements [1]. Phosphorene nanoribbons, nanotubes and fullerene-like structures, just as graphene nanoribbons, carbon nanotubes and fullerenes are currently the subject of intensive research [1,2]. For example, comprehensive theoretical studies have been carried out to explore the electronic and mechanical properties of phosphorene nanoribbons. It was found that phosphorene nanoribbons can be either metals (with zig-zag edges) or semiconductors (with arm-chair edges). The band gap of arm-chair nanoribbons can be tuned by strain [23], electric field [54–56], chemical functionalization [49,57,58] and edge passivation [54,59]. Although not as mechanically tough as graphene, boron-nitride or TMD-based nanoribbons as might be hoped, the in-plane stiffness of phosphorene nanoribbons is still much larger than that of many commonly used materials [34–36].

Phosphorene nanotubes (PNTs) and fullerene-like nanoparticles are less explored. For example, research has been conducted to study the electronic properties of PNTs [54,60]. It was found that PNTs are direct band gap semiconductors. The electronic properties of PNTs can be tuned from semiconducting to metallic by applying strain or electric field [54]. PNTs can be formed by deforming monolayer phosphorene at substantial strain energy cost. Guan et al. [61] suggested that this energy cost can be avoided by forming faceted nanotubes using a variety of stable planar phases of phosphorene, which can be connected at essentially zero energy cost. This option to mix different phases offers exceptional richness in PNT forms and their associated properties [61].

While some of the elastic properties of phosphorene nanotubes have been studied [60], their deformation and failure mechanisms at large strain, however, are mostly unexplored. This is in strong contrast with the significant amount of works performed on other 2D materials, such as graphene [62–66], boron-nitride [67–70] and dichalcogenides nanotubes [13,71–73]. The behavior of PNTs under straining, in particular, their deformation and failure mechanisms and their critical strain and stress, must have been studied comprehensively in order to use PNTs in future nano-mechanical devices.

Our goal is to investigate the structural properties, deformation and failure of armchair and zigzag PNTs under large uniaxial tensile straining. Our objective is to find the answers for the following questions: What are the structures of PNTs? What are the fracture mechanisms of PNTs? Does the puckered structure of phosphorene profoundly affect their deformation and failure? Is there an evident mechanical anisotropy in PNTs? Are the failure strain and strength dependent on the geometry and the diameter of PNTs? To answer these questions, we carry out density functional tight binding calculations, focusing on the two primary geometries of nanotubes: armchair (AC) and zigzag (ZZ) phosphorene nanotubes.

## 2. Computational Model

Single-wall AC and ZZ PNTs can be constructed by rolling up phosphorene monolayer along the AC and ZZ directions (see Figure 1). The constructed geometries can be described by the number of phosphorene unit cell aligned along the nanotube circumference.

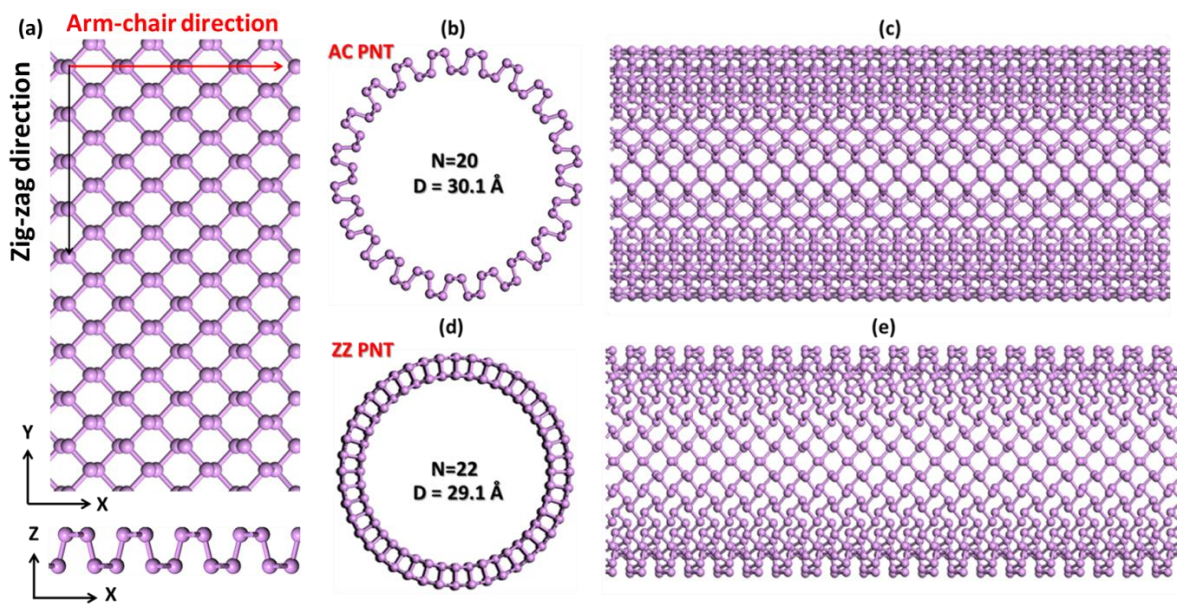

**Figure 1:** (a) Upper panel (top view): phosphorene monolayer with the AC (along the X-axis) and ZZ (along the Y-axis) directions indicated. Lower panel (side view): puckered structure of phosphorene. (b, c) AC phosphorene nanotube with diameter D = 30.1 Å (N=20, where N is the number of unit cell aligned along the nanotube circumference): top (b) and side (c) views; (d, e) ZZ phosphorene nanotube with diameter D =29.1 Å (N=22): top view (d) and side view (e).

We apply tight-binding (TB) technique [74–77] to examine the deformation and failure of PNTs under uniaxial tensile strain. The TB technique has a special place among the computational techniques accessible for nanoscale modelling of materials. On one hand, the density function theory (DFT) method, which has been extensively used to explore nanoscale phosphorene-based systems, is very precise, but computationally demanding. As a result, the DFT calculations are not feasible for large or even intermediate scale systems. On the other hand, molecular dynamics (MD) simulations cannot be used either since trustworthy, highly-reliable and widely accepted interatomic potential for phosphorene is yet not available. In this situation, empirical tight-binding technique, situated in between DFT and MD in terms of computational cost and accuracy, is an ideal approach to deal with the size problem.

In our simulations of PNTs, we apply density functional tight-binding (DFTB) method [78], which appropriately combines the accuracy of DFT with the computational effectiveness of TB. The DFTB is derived from DFT but uses empirical approximations to increase the computational efficiency, while preserving accuracy [79–81]. The substitution of the many-body Hamiltonian of DFT with a parameterized Hamiltonian matrix is the key approximation in the TB method [82]. In the DFTB approach, quasi-atomic wave functions, represented in terms of Slater-Koster type orbitals [74] and spherical harmonics [83], are applied to model the electron density. The orbital basis is created and fitted using DFT calculations, and then applied to calculate the Hamiltonian and overlap matrix elements [78]. These matrix elements do not completely define the total energy of the system. The residual part of the total energy is added as a short-range repulsive term, which can be represented in terms of pair wise potentials between atoms. The pair wise potentials are obtained via a fitting routine. Besides the short-range and electronic repulsive terms, the Kohn-Sham energy contains the dispersion interactions (van der Waals forces) and Coulomb interactions [78]. At a long distance, the second term describes long-range electrostatic interactions between two point charges and also includes self-interaction contributions of a given atom (if the charges are positioned at the same atom) [84].

Furthermore, one uses self-consistent charge (SCC) technique in DFTB to improve the description of atom bonding [84]. Due to the SCC extension, the DFTB can be effectively applied to problems where shortages in the non-SCC standard TB method come to be evident [78]. Combining almost quantum mechanical precision and computational efficiency similar to molecular dynamics methods, the DFTB comes up with exceptional opportunities to explore nano-systems inclosing a few hundreds of atoms. For instance, structural, mechanical and electronic properties of phosphorene monolayer, nanoribbons, and nanotubes were investigated earlier by using DFTB [26,55,60,85–88].

We note, that the mechanical properties of nanotube obtained using different methods such as DFT, DFTB, and MD are not the same [89]. While studying the mechanical properties of carbon nanotubes Zhang et al. [89] showed that the most accurate results are offered by the most computationally expensive and slow DFT calculations, while much faster MD simulations produce less accurate results, since they neglect electronic degrees of freedom, as well as all quantum effects. The application DFTB yields a suitable balance between MD efficiency and DFT precision.

Following our previous approach in DFTB simulations [88,90] optimized the unit cell of phosphorene found by DFT method [20]. Using the DFTB optimized unit cell, we constructed a phosphorene monolayer and rolled it up into the AC (see Figure 1(a)) and ZZ (see Figure 1(b)) PNTs. In what follows, we investigate the mechanical response of PNTs with different diameters (see Figure 2) exposed to uniaxial tensile strain. By varying the number of unit cells along the nanotube circumference from $N=8$ to $N=22$, we constructed AC PNTs with the diameter in the range from $D=13.3$ Å to $D=32.8$ Å, and ZZ

PNTs from D=13.2 Å to D=31.1 Å. The length of the AC and ZZ PNTs was chosen to be L=14.8 Å and L=16.4Å, respectively. In order to check the effect of nanotube length, we doubled the length and found that the obtained results for the shorter nanotubes did not deviate significantly from the longer ones. The total number of atoms used in our DFTB simulations varied in the range between ~100 to ~350.

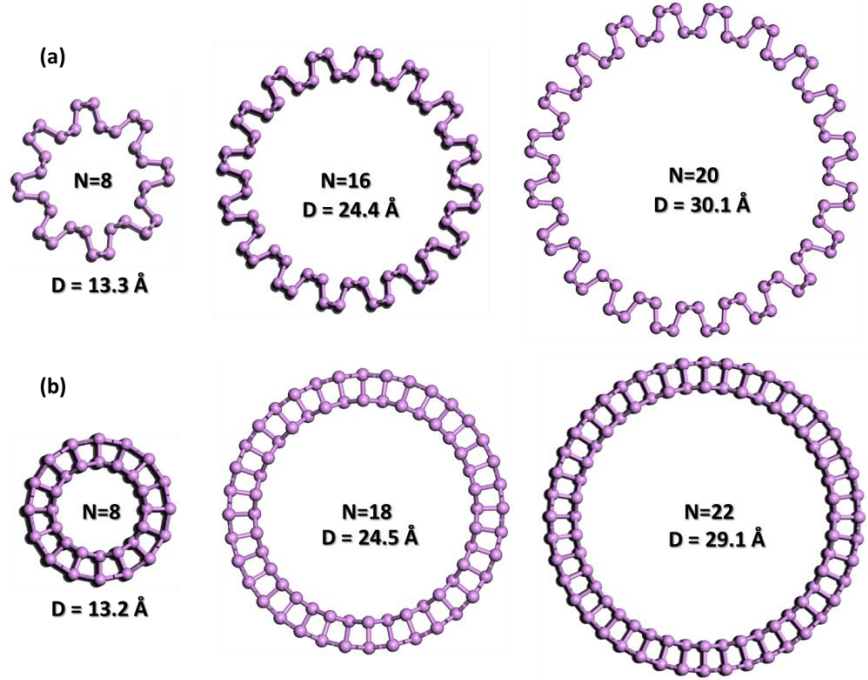

Figure 2: (a) DFTB optimized AC PNTs with different diameters: D = 13.25 Å (N=8), D = 24.44 Å (N=16) and D = 30.05 Å (N=20). (b) DFTB optimized ZZ PNTs with different diameters: D = 13.22 Å (N=8), D = 22.26 Å (N=18) and D = 29.06 Å (N=22). Here, N is the number of unit cell along the tube circumference.

In our DFTB simulations, we enforced periodic boundary conditions in all directions. The uniaxial strain was applied along the nanotube axis (X-axis). To avoid the self-interaction of PNTs due to periodic boundary conditions along the Y and Z axes, a vacuum slab was added in the radial directions perpendicular to the tube axis. The width of the vacuum slab was chosen to be 20 Å. The k-point set for the Brillouin-zone integration was chosen by using the Monkhorst-Pack method [91–93].The Monkhorst-Pack grid [91,92] with an 8x2x2 sampling set was adapted for Brillouin-zone integration. Following the DFTB studies for PNTs [60], the s- and p-orbitals were specified for every P-atom. The Slater-Koster files [74] for phosphor atoms were taken from 'MATSCI' set [94,95].

We applied a uniform uniaxial tensile strain to the PNTs quasi-statically along the tube axis (X-axis) at zero temperature. The tensile strain was increased gradually by a small step of $\delta\varepsilon$=0.01. Consequently, the initial configuration of the PNTs was relaxed by minimizing the total energy of the system using conjugate gradient method. The self-consistent charge calculations were carried out at each step of the energy minimization. We estimated a nominal stress, as explained in [20], by calculating the force per unit cross-sectional of effective area of the PNT, which is obtained as the difference between the areas of the outer and the inner circles, outlined by the outer and the inner radii of the PNT (see Figure 2). The volume of the nanotube is determined as a product of its length and the effective area. The nanotube thickness is defined as the difference between the outer and the inner radii of the PNT.

We note that in our DFTB simulations, we do not consider the end effects, since PNTs are modeled as of infinitely long nanotubes, stretched uniformly along the axial direction. However, it is expected that the end effects can largely affect their mechanical and electronic properties, just as in carbon nanotubes [96]. The effect of the nanotube ends on the mechanical properties of PNTs is the subject of our forthcoming research.

## 3. Results and Discussion

### 3.1 Geometry optimization

As the first step, we optimized the geometry of the constructed PNTs by minimizing their total potential energy. After the relaxation of the PNT structures, we obtained the tube diameters and lengths at zero strain. Both the bond lengths and bond angles in the phosphorene sheet have to be changed to accommodate for the tube curvature. In monolayer phosphorene, P-atoms are linked in a ring-like structures, consisting of six P-atoms (see Figure 1), similar to the hexagonal graphene rings. Yet, unlike graphene, phosphorene is not perfectly flat; instead, due to the sp$^3$ hybridization, the P-atoms form a nonplanar puckered (folded) accordion-like structure, resembling a hexagonal atomic plane under compression. This puckered structure maximizes the distance between the lone electron pairs located at each P-atom [54,60]. Each phosphorene atom is connected to the three nearest neighbors by covalent bonds as shown in Figure 3(a). The two of its neighbors are in the same [XY] plane, and together they form an angle of θ=96.16°. The third neighbor is located in the adjacent plane at the angle γ=102.42° (see Figure 3(a)). The bond length between the adjacent in-plane atoms (denoted as the **A**-bond in Figure 3(a)) is somewhat shorter than the bond length between the nearest atoms located in different planes (denoted as the **B**-bond in Figure 3(a)). The lengths of the A-bonds located in the lower or upper plane are the same.

When a phosphorene sheet is rolled into a tube, the symmetry between the lower and upper planes and the corresponding equivalence of the bond length, are broken. The bond length of the A-bonds in the inner shell (denoted as $A_I$ bonds in Figure 3, where the subscript I denotes the inner tube shell) shortens; while the bond length of the A-bonds in the outer shell (denoted as $A_O$ bonds in Figure 3, where the subscript *O* denotes the outer tube shell) lengthens. Apparently, the degree of the bond length contraction or elongation depends on the tube diameter, similar to carbon nanotubes [97–101]. The differences between the inequivalent bonds increase with the decrease in the tube diameter. The bond lengths of the $A_I$, $A_O$ and B-bonds as a function of tube diameter at zero strain are shown in Figure 4(a) for AC and Figure 4(b) for ZZ PNTs. The differences in the bond length between the PNT and phosphorene monolayer are clearly visible. As expected, the bond length of the inner $A_I$ bonds significantly decreases, while the bond length of the outer $A_O$ bonds increases with the tube diameter. The bond length contraction for the inner $A_I$ bonds is more evident for AC PNTs than for ZZ PNTs. The difference between the $A_I$ and $A_O$ bonds also increases with the decrease in the tube diameter. Both the effects of bond length contraction and elongation can be understood in terms of the curvature-induced re-hybridization of P-P orbitals [60].

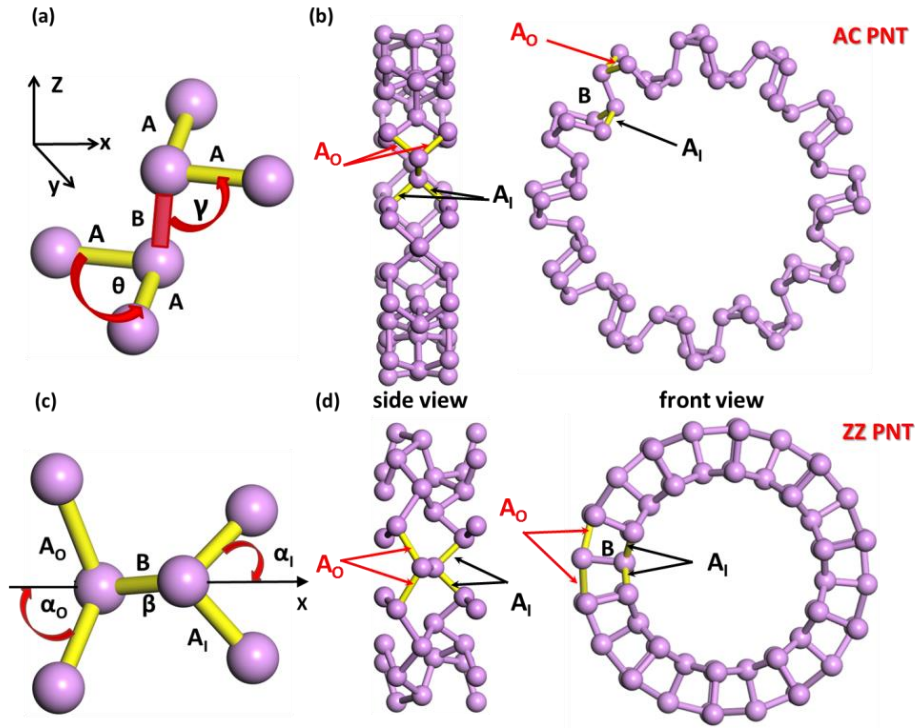

Figure 3: (a) Two types of P-P bonds in puckered monolayer phosphorene: the B-bond marked by red and A-bonds marked by yellow. (b), (d) A-bonds situated in the outer ($A_O$-bonds, red arrows) and inner shell ($A_I$-bonds, black arrows) of AC (b) and ZZ (d) PNTs. Side and front views are used to illustrate the positions of the bonds. (c) Angles $\alpha_I$, $\alpha_O$, $\beta$ between $A_I$-bond, $A_O$-bond, B-bond and the stretching direction (X-axis), respectively.

The length of the B-bonds, which connect to the $A_I$ and $A_O$ bonds, also depends on the tube diameter. In the case of AC PNTs, the length of B-bonds, which are oriented at a right angle to the direction of the applied strain, is noticeably larger than that of phosphorene monolayer. It decreases monotonously with the tube diameter, and finally converges to that of phosphorene monolayer. In contrast, for ZZ PNTs, the length of the B-bonds at smaller tube diameters is shorter than in phosphorene monolayer. It increases evenly with the tube diameter, and finally converges to bond length of phosphorene monolayer (see Figure 4).

Besides the change in the bond length, there is a slight bond reorientation. In what follows, we examined the angles between the stretching direction and nanotube bonds. Three angles, $\alpha_I$, $\alpha_O$, and $\beta$, which are the angles between the $A_I$-, $A_O$- and B-bonds and stretching direction (X-axis), are exemplified in 3(c). As shown in Figure 4(c), the $A_I$ and $A_O$ bonds in AC PNTs are arranged at an acute angle ($\alpha_I, \alpha_O \approx 45°$), while the B-bonds are approximately normal to the stretching direction ($\beta \approx 90°$). The values of these angles are unaffected by the AC tube diameter. On the contrary, the $A_I$ and $A_O$ bonds of ZZ PNTs are arranged at two different angles ($\alpha_I \approx 45°$ and $\alpha_O \approx 55°$), while the B-bonds are aligned at an acute angle (β≈75°) to the direction of applied strain (see Figure 4(d)). For ZZ PNTs, the values of these angles depend on the tube diameter. The value of the $α_O$ angle decreases as the ZZ tube diameter increases.

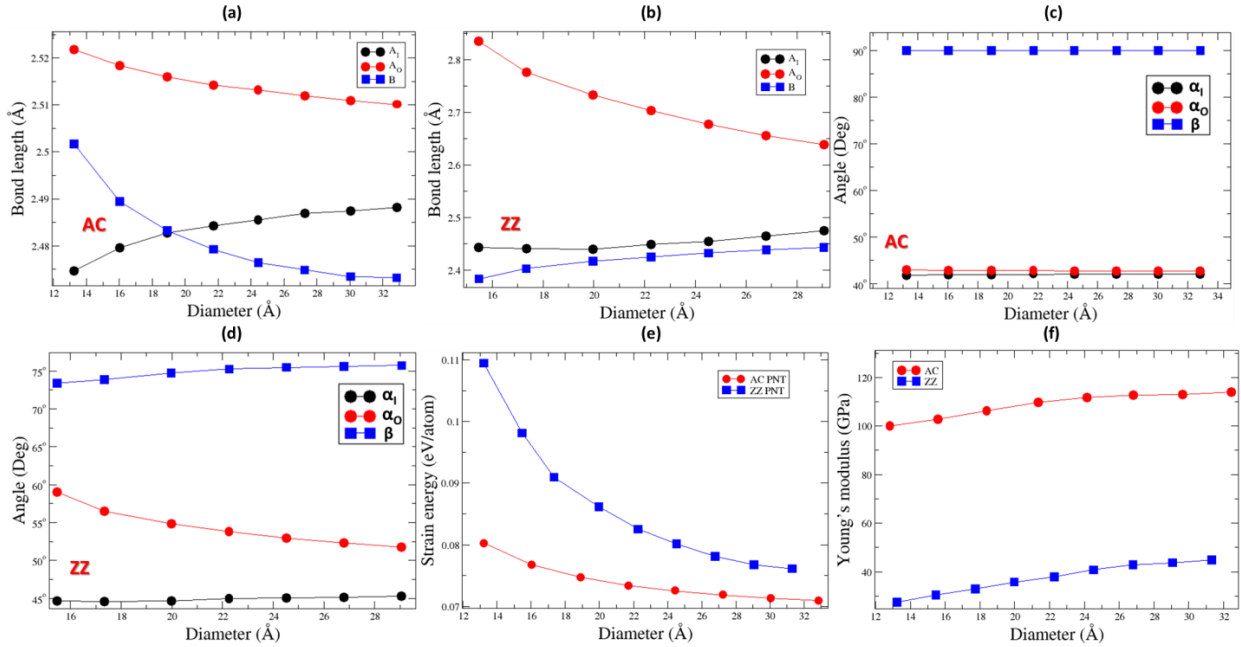

Figure 4: (a) Bond length of $A_I$ (black circles), $A_O$ (red circles) and B (blue squares) bonds of AC PNT are plotted vs nanotube diameter. (b) Bond length of $A_I$ (black circles), $A_O$ (red circles) and B (blue squares) bonds of ZZ PNT are plotted against nanotube diameter. (c) Angle between $A_I$ (black circles), $A_O$ (red circles) and B (blue squares) bonds and the stretching direction for AC PNT as a function of nanotube diameter. (d) Angle between $A_I$ (black circles), $A_O$ (red circles) and B (blue squares) bonds and the stretching direction for ZZ PNT as a function of nanotube diameter. (e) Strain energy of AC (red circles) and ZZ (blue squares) PNTs as a function of tube diameter. (f) Young's modulus of AC (red circles) and ZZ (blue squares) PNTs as a function of tube diameter.

Due to the variations in the bond lengths and angles, the actual tube diameters and lengths are slightly different from the ideal ones obtained by rolling a phosphorene sheet, and the relaxation effects are more pronounced for PNTs with smaller diameters, and become negligible when the tube diameter is large, similar to carbon nanotubes [97,98].

## 3.2 Flexural rigidity and Young's modulus

Next, we calculated the strain energy per atom associated with rolling up a phosphorene sheet in a tubular structure. The strain energy, $E_s$, was calculated as a difference between energy (per atom) of the tubular structure and phosphorene monolayer: $E_s = \frac{E_{PNT}}{N} - E_{at}$, where $E_{PNT}$ is the total potential energy of the nanotube, N is the number of the nanotube atoms and $E_{at}$ is the energy per atom in phosphorene monolayer. The strain energy as a function of tube diameter is shown in Figure 4(e). As can be seen from Figure 4(e), the strain energy per atom for ZZ PNTs is larger than that for AC PNTs. The difference is particularly noticeable for nanotubes with small diameters, and it decreases rapidly when the tube diameter increases. The lower strain energy per atom of AC PNTs indicates that they are energetically more favorable and stable than ZZ PNTs, in agreement with previous DFT calculations [54,60].

As expected from the linear elasticity theory [102–105], the strain energy varies as $E_S = \frac{\delta}{D^2}$ with the tube diameter, $D$. Using data in Figure 4(e), we extracted the flexural (bending) rigidity, $\delta$. It was found that the flexural rigidity for AC PNT is $\delta_{AC}$=0.019 (eV nm$^2$/atom), which is almost three times smaller than that for ZZ PNT $\delta_{ZZ}$=0.071 (eV nm$^2$/atom).

Additionally, we obtained the Young's modulus, $Y$, for AC and ZZ PNTs, which is related to the second derivative of the total potential energy, $E_{PNT}$, with respect to the strain (calculated at zero strain). More specifically,

$$Y = \frac{1}{V_0}\left(\frac{\partial^2 E_{PNT}}{\partial \varepsilon^2}\right)_{\varepsilon=0}$$

where $V_0$ is the nanotube volume at zero strain. The second derivative of the total potential energy was calculated by subjecting PNTs to small uniaxial compressive and tensile strains (ε ≤ 3%), and subsequent optimization of the atomic positions at each strain. The total potential energies for these small deformations were fitted to a second order polynomial, from which the second derivative at zero strain was found.

The obtained Young's modulus for AC and ZZ PNTs as a function of tube diameter is shown in Figure 4(f). It is seen that the Young's modulus of AC PNTs is larger than that of ZZ PNTs in the examined range of the tube diameters. The value of Young's modulus for the ZZ PNT with the largest diameter is Y≈40 GPa, while for the AC PNT of the similar diameter is Y≈115 GPa, which is almost three times larger than that of the ZZ PNT. Another interesting observation is that the Young's modulus slightly depends on tube diameter. It gradually increases with the tube diameter and converges to a diameter-independent constant value (see Figure 4(f)). The effect of the tube diameter on the elastic properties of PNTs is only significant for small diameter PNTs (D ≤ 20Å, see Figure 4(f)). The similar effect of tube softening with decreasing diameter was also found for carbon and boron-nitride nanotubes with small diameters [97,104,106–108]. This minor reduction in the Young's modulus at small diameters can be ascribed to the minor weakening of P-P bonds due to the tube curvature. As can be seen in Figure 4 (a, b), the P-P bond (namely, the $A_I$ and $A_O$ bonds for AC PNTs and the B-bonds for ZZ PNTs), stretched along the direction of applied tensile strain, are initially pre-compressed due to the tube curvature. Therefore, when the uniaxial tensile strain is applied, the length of these bonds first reverts to its equilibrium bulk value, and then starts to increase. Hence, the Young's modulus reduces as the tube diameter decreases.

### 3.3 Poisson's ratios
We calculated the radial Poisson's ratio, $\nu_D$, which is defined as the ratio of the transverse contraction strain $\frac{\Delta D}{D}$ to the longitudinal extension strain $\frac{\Delta L}{L}$ of PNTs, and can be defined as:

$$\frac{\Delta D}{D} = -\nu_D \frac{\Delta L}{L},$$

where $D$ is the tube diameter and $L$ is the tube length. The Poisson's ratio for the tube thickness, $\nu_t$, was calculated in the same way by replacing the diameter $D$ with the tube thickness, $t$, which is defined as the difference between the outer and inner tube radii. The Poisson's ratios $\nu_D$ and $\nu_t$ are both positive since the elongation of PNTs along the axial direction reduces the diameter and thickness (see Figure 5).

The diameter of AC and ZZ PNTs decreases in a different way with applied tensile strain. As can be seen from Figure 5(a, b), AC PNTs contract in the transverse direction much more than ZZ PNTs. This distinction is related to the geometrical structure of PNTs. The puckered structure of AC PNTs, where the phosphorene puckers are aligned along the tube circumference, accounts for the substantial compressibility in the radial direction (see Figure 5(a)). A pucker in phosphorene resembles a hinge, i.e. a jointed flexible device that allows the turning of a part on a stationary frame. In these puckers, the B-bonds, which are oriented along the radial direction and connect to adjacent $A_I$ and $A_O$ bonds (see Figure 3), can pivot around. Thus, the rotation of B-bonds, together with their minor contraction, leads to a considerable compression of AC PNTs in the radial direction. In ZZ PNTs, where the phosphorene puckers are aligned along the tube axis, the diameter reduction in the radial direction is due to the compression of the $A_I$ and $A_O$ bonds. Since the bond compression costs a significant amount of energy, the contraction in the transverse direction of ZZ PNTs is smaller than that of AC PNTs.

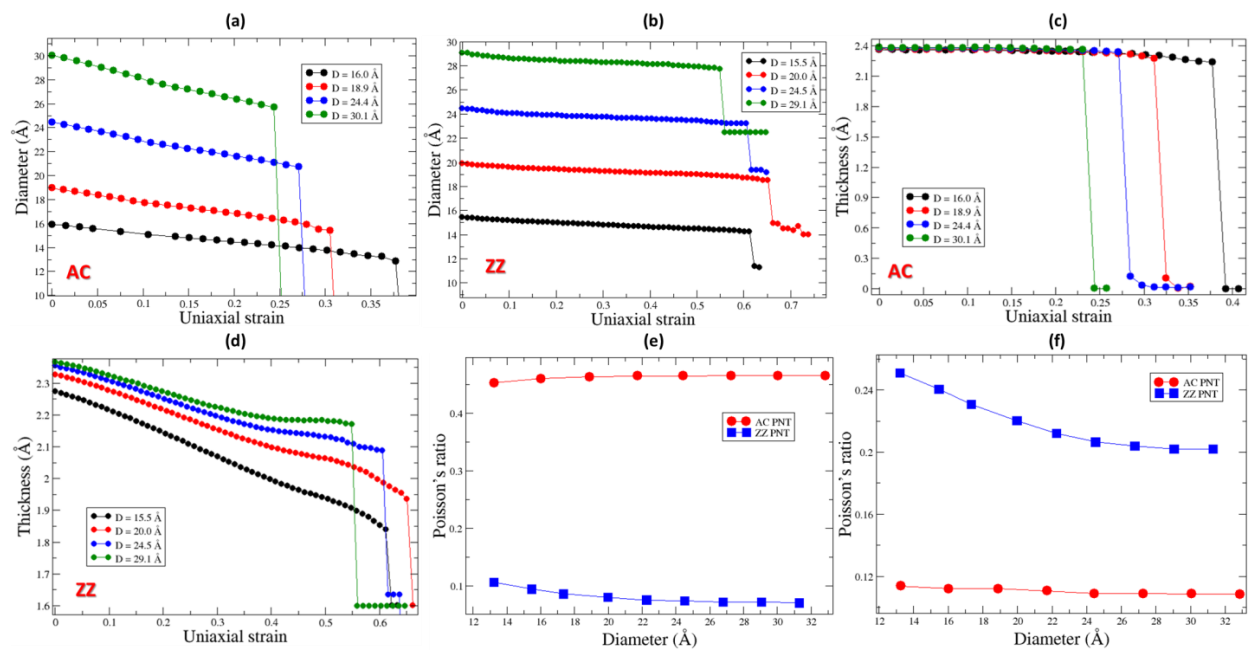

Figure 5: (a) Diameter of AC PNTs as a function of uniaxial tensile strain. Initial diameters (measured at zero strain) of AC PNTs are D=16.0 Å (black circles), D=18.9 Å (red circles), D=24.4 Å (blue circles) and D=30.1 Å (green circles), respectively. (b) Diameter of ZZ PNTs is plotted vs. applied uniaxial tensile strain. Initial diameters (measured at zero strain) are D=15.5 Å (black circles), D=20.0 Å (red circles), D=24.5 Å (blue circles) and D=29.1 Å (green circles). (c) Thickness of AC PNTs as a function of applied tensile strain. Initial diameters are D=16.0 Å (black circles), D=18.9 Å (red circles), D=24.4 Å (blue circles) and D=30.1 Å (green circles). (d) Thickness of ZZ PNTs as a function of applied tensile strain. Initial diameters are D=15.5 Å (black circles), D=20.0 Å (red circles), D=24.5 Å (blue circles) and D=29.1 Å (green circles). (e, f) Poisson's ratio for contraction in the transverse direction of tube diameter (e) and thickness (f) of AC (red circles) and ZZ (blue squares) PNTs due to tensile strain applied in the longitudinal direction.

The tube thickness changes with applied tensile strain in an opposite way. The thickness of AC PNTs with various diameters changes insignificantly with the applied tensile strain (see Figure 5(c)). On the other hand, the thickness of ZZ PNTs decreases rapidly with the applied strain as shown in see Figure 5(d), since the phosphorene puckers (which can be easily squeezed due to bond rotation) located along the nanotube circumference. In addition, there is also an apparent dependence of the tube thickness on the tube diameter. As can be seen in Figure 5(c), the smaller is the ZZ PNT diameter, the larger is the decrease in its thickness.

We calculated both the radial (see Figure 5(e)) and thickness (see Figure 5(f)) Poisson's ratios as a function of diameter for AC and ZZ PNTs. The radial Poisson's ratio for AC PNTs ($\nu_D$ = 0.47) is almost seven times larger than that for ZZ PNTs ($\nu_D$ = 0.07). Similar to carbon nanotubes, the radial Poisson's ratio moderately depends on the tube diameter (a slight reduction for small radii) [97,104,106,108]. The value of the radial Poisson's ratio converges to a constant at the large diameters (see Figure 5(f)). The thickness Poisson's ratio for AC PNTs is a constant ($\nu_t$=0.11), independent of tube diameter; while the thickness Poisson's ratio of ZZ PNTs decreases from $\nu_t$=0.25 and gradually converges to $\nu_t$=0.21 as the tube diameter increases.

We summarize the calculated values for the elastic moduli, flexural rigidity and Poisson's ratios of PNTs in Table 1. Since the elastic moduli and Poisson's ratios only slightly or moderately depend on the tube diameter, we use the converged values for the large tube diameters.

Table 1: Elastic modulus, flexural rigidity and Poisson's ratios of phosphorene nanotubes

| Nanotube type | Young's modulus (GPa) | Flexural rigidity (eV nm$^2$/atom) | Poisson's ratio (radial) | Poisson's ratio (thickness) |
|---|---|---|---|---|
| AC | 114.2 | 0.019 | 0.47 | 0.11 |
| ZZ | 49.2 | 0.071 | 0.07 | 0.21 |

### 3.4 ZZ PNTs: tensile deformation and failure pattern

In the following, we investigate the deformation and failure of ZZ PNTs at large strains. The strain energy of ZZ PNTs is shown in Figure 6(a) for a selected set of ZZ PNTs with different diameters. We also calculated the stress as a function of tensile strain applied for the same set of ZZ PNTs (see Figure 6(b)).

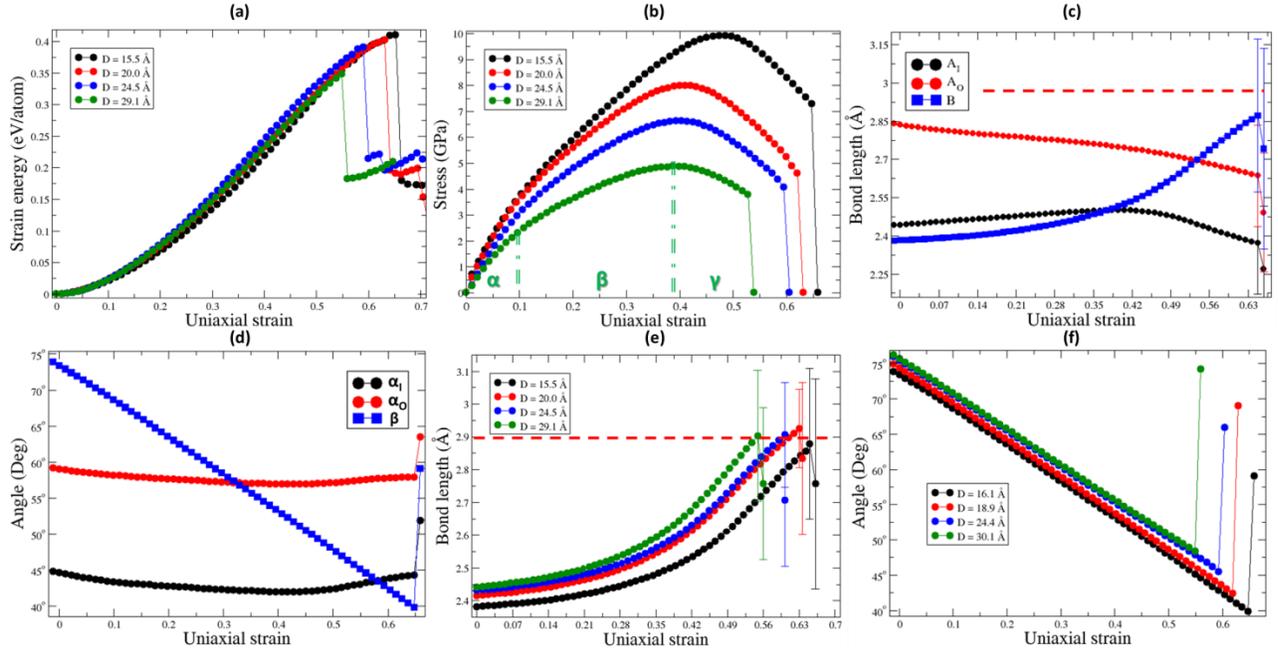

Figure 6: (a) Strain energy (per atom) plotted vs. uniaxial tensile strain for ZZ PNTs of different diameters: D=15.5 Å (black circles), D=20.0 Å (red circles), D=24.5 Å (blue circles) and D=29.1 Å (green circles). (b) Stress-strain curves for ZZ PNTs of different diameters: D=15.5 Å (black circles), D=20.0 Å (red circles), D=24.5 Å (blue circles) and D=29.1 Å (green circles). Three phases of ZZ PNT deformation α, β, and γ are indicated. (c) Average bond length of $A_I$ (black circles), $A_O$ (red circles) and B (blue squares) bonds of ZZ PNT (D=24.5 Å) as a function of applied tensile strain. Error bars indicate the bond length variation at the failure strain. The red dashed line specifies the maximally allowed bond length. (d) Average angles between the $A_I$ (black circles), $A_O$ (red circles) and B (blue squares) bonds and stretching direction for ZZ PNT (D=24.5 Å) are plotted against applied tensile strain. (e) Average bond length of the B-bonds and (f) angle between B-bonds and stretching direction are plotted vs. applied tensile strain for ZZ PNTs with D=15.5 Å (black circles), D=20.0 Å (red circles), D=24.5 Å (blue circles) and D=29.1 Å (green circles).

There are three characteristic phases (marked as α, β, γ in Figure 6(b)) in the tensile deformation of ZZ PNTs. At a small strain ($\varepsilon \lesssim 0.1$), the linear elastic phase (α-phase) was identified; when the tensile strain surpassed $\varepsilon \approx 0.1$, the bond rotation phase was observed (β-phase), where the folded structure of phosphorene nanotubes was unfolded via the B-bond rotation. As a final point, when the tensile strain exceeded $\varepsilon \gtrsim 0.4$, the direct B-bond stretching phase followed (γ-phase), which was ended by the B-bond breaking at the failure strain of $\varepsilon_{cr} \approx 0.6$.

At the α-phase, we found that the potential energy increases non-linearly as a quadratic function of applied tensile strain, while the stress increases linearly (see Figure 6(a, b)). This phase completes when the uniaxial strain approaches $\varepsilon \approx 0.1$. Similar to phosphorene nanoribbons[90], the end of the linear elastic phase is associated with the puckered structure of phosphorene: each P-atom is strongly connected to its nearest neighboring atoms by covalent bonds. Moreover, it interacts with its next nearest neighbors located in the neighboring phosphorene puckers (see Figure 1(a)). The pucker-pucker distance falls within the radius of inter-atomic interaction[90] at zero strain, however, upon tensile stretching, the distance between the two atoms from the adjacent puckers increases. This range of pucker-pucker inter-atomic interaction is exceeded at the strain $\varepsilon \approx 0.1$. Therefore, above $\varepsilon \approx 0.1$, the interaction between the pucker atoms is negligible, and the linear elastic phase terminates.

Subsequently, the elongation of ZZ PNTs is achieved via the rotation of B-bonds (see Figure 6(b)). The change in the length of B-bonds in the β-phase is negligible (see Figure 6(c, e)). The B-bond rotation unfolds the puckered structure of ZZ PNTs. The rotation angle, defined as an angle between a B-bond and the direction of applied strain is plotted in Figure 6(d): it decreases linearly with applied strain from β≈75° to β≈40°~50° at the failure strain, depending on the tube diameter (see Figure 6(f)).

When the applied strain is larger than ε≈0.4, the B-bond rotation is replaced by the B-bond stretching (see the γ-phase in Figure 6(b)), in which the B-bond length increases steeply with applied tensile strain (see **Error! Reference source not found.** and Figure 7(b)). At the same time, the bond length of $A_I$ and $A_O$ bonds decreases. As soon as the critical tensile strain $ε_{cr}$≈0.6 is reached, the B-bonds break along the entire length of the nanotube (see Figure 7(c)), releasing the accumulated strain energy. A P-P bond ruptures when its length exceeds the maximal bond length, which is about $l_{max}$=2.9 Å [88,90]. The pattern of the ruptured bonds at $ε_{cr}$≈0.6 is shown in Figure 7(c). It should be noted that only a fraction of the B-bonds rupture at the critical strain. The bond length variation at the failure strain is indicated by the error bars in Figure 6(c).

These three phases of deformation are universal for ZZ PNTs with different diameters. As shown in Figure 6(c), the B-bond rotation and elongation phases are the same for all ZZ PNTs with different diameters, although the particular values of tensile strain at which one phase is replaced by another phase and at which ZZ PNTs fail depend slightly on the tube diameter.

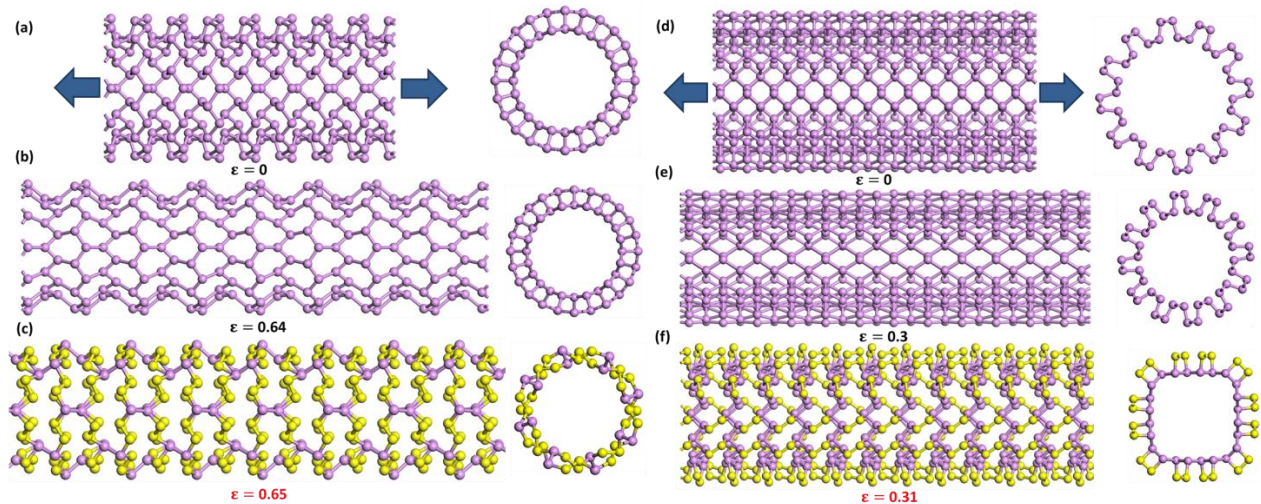

**Figure 7:** (a-c) Deformation and failure of ZZ PNTs under applied uniaxial tensile strain (side and front views). Sequence of snapshots of ZZ PNT with D=24.5 Å at different strains: (a) ε=0 (reference configuration), (b) ε=0.64 (c), and $ε_{cr}$=0.65 (critical strain). P atoms that are connected by less than three P-P bonds are marked by yellow. (d - f) Deformation and failure of AC PNTs subjected to uniaxial tensile strain (side and front views). Sequence of snapshots of AC PNT with D= 18.6 Å at different tensile strains: (d) ε=0 (reference configuration), (e) ε=0.3, and (f) ε=0.31 (critical strain).

## 3.5 AC PNTs: tensile deformation and failure mechanism

Next, we studied the deformation of AC PNTs under uniaxial tensile strain. The calculated strain energy and stress are shown in Figure 8(a, b). The stress-strain curves for AC PNTs reveal a very different characteristic of tube deformation: a rapid increase in the strain energy, which culminates at the failure of AC PNTs when the length of the strained bonds exceeds the maximally allowed limit. For AC PNTs, unlike to ZZ PNTs, the applied strain causes the direct straining of the $A_I$ and $A_O$ bonds from the beginning (see Figure 8(c)). Their bond length increases as a linear function of the uniaxial strain regardless the tube diameter (see Figure 8(e)). The orientation of the $A_I$ and $A_O$ bonds also changes accordingly, as they pivot to align along the direction of stretching (see Figure 8(d)). The change in the rotation angles $\alpha_I$ and $\alpha_O$ (see Figure 3(c)) is about $\Delta\alpha_I=\Delta\alpha_O\approx10°$ (see Figure 8(f)). The angles $\alpha_I$, and $\alpha_O$ decrease in the same way irrespective of the tube diameter as shown in Figure 8(d). At the same time, the length of the B-bonds shortens, causing AC PNTs to contract in the radial direction. The orientation of the B-bonds (oriented perpendicularly to the direction of applied strain) remains constant until the failure strain is reached (see Figure 8 (d)).

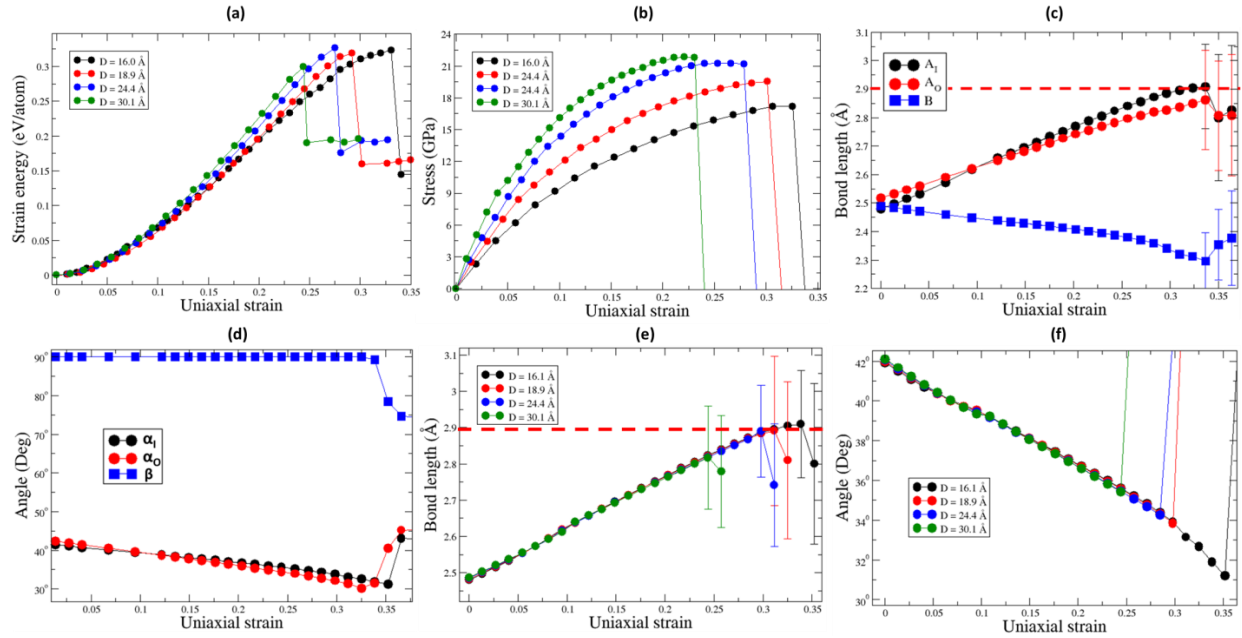

Figure 8: (a) Strain energy per atom as a function of applied uniaxial tensile strain for AC PNTs with different diameters: D=16.0 Å (black circles), D=18.9 Å (red circles), D=24.4 Å (blue circles) and D=30.1 Å (green circles). (b) Stress-strain curves of AC PNTs with different diameters: D=16.0 Å (black circles), D=18.9 Å (red circles), D=24.4 Å (blue circles) and D=30.1 Å (green circles). (c) Average bond length of $A_I$ (black circles), $A_O$ (red circles) and B (blue squares) bonds of AC PNT (D=18.9.5 Å) as a function of applied tensile strain. Error bars indicate the bond length variation at the critical strain. The red dashed line specifies the maximally allowed bond length. (d) Average angle between $A_I$ (black circles), $A_O$ (red circles) and B (blue squares) bonds and the stretching direction for AC PNT (D=18.9 Å) vs. applied tensile strain. (e) Average bond length of the B-bonds and (f) angle between B-bonds and stretching direction are plotted against the applied tensile strain for AC PNTs with D=16.1 Å (black circles), D=18.9 Å (red circles), D=24.4 Å (blue circles) and D=30.1 Å (green circles).

In order to understand the failure mechanism of AC PNTs, the AC PNT with diameter D=18.6 Å was examined. The snapshots for the failure process are shown in Figure 7(d-f). It is seen that near the failure strain, the $A_I$ and $A_O$ bonds are stretched to their limit (see **Error! Reference source not found.**(e)), and at the critical strain $\varepsilon_{cr}\approx0.31$, a fraction of these strained bonds rupture. The substantial bond length variation at the critical strain is indicated by error bars in Figure 8(e). The ruptured $A_I$ and $A_O$ bonds are uniformly distributed along the AC nanotubes. Due to the radial contraction of AC PNTs in the transverse direction, the distance between the neighboring puckers, located along the nanotube circumference, decreases substantially. As a result the adjacent puckers are drawn close enough to form bonds between the atoms located in the inner tube shell at $\varepsilon_{cr}\approx 0.31$. The bond rupture at the failure strain is accompanied by the formation of the new bonds and irreversible rearrangement of the old bonds between the atoms of the inner and outer tube shells (see **Error! Reference source not found.**(f)).

## 3.6 Critical strain and stress

Our simulations indicate that the failure strain for AC PNTs is in the range from $\varepsilon\approx 0.25$ to $\varepsilon\approx 0.4$, and for ZZ PNTs it is between $\varepsilon\approx 0.55$ and $\varepsilon\approx 0.66$ (see Figure 9). Thus, the failure strain of ZZ PNTs is almost two times larger than that of AC PNTs. In ZZ PNTs, the applied strain causes the rotation and straining of the B-bonds connecting P atoms between the adjacent tube shells, while in AC PNTs, the $A_I$ and $A_O$ bonds linking atoms within the inner and outer tube shells are the most strained. Evidently, the noticeable variances in the mechanical properties and failure behavior of the ZZ and AC PNTs originate from the difference in the nanotube geometry. From the obtained stress-strain curves for ZZ PNTs (see Figure 6(b)) and AC PNTs (see Figure 8(b)), we calculated the failure strain and failure stress as a function of nanotube diameter, which are shown in Figure 9. The failure strain as a function of tube diameter for AC (red circles) and ZZ (blue squares) PNTs is plotted in Figure 9(a). For AC PNTs, the failure strain decreases steadily from $\varepsilon_{cr}\approx 0.4$ to $\varepsilon_{cr}\approx 0.25$ with the tube diameter within the range examined in our DFTB simulations. This decrease in the critical strain with the tube diameter is related to the fact that AC PNTs with smaller diameters are partially reinforced by an increase in the magnitude of interactions between the adjacent puckers. This is because the distance between the adjacent puckers is shorter for AC PNTs with smaller diameters due to their contraction in the transverse (radial) direction under applied axial strain. The increase in the magnitude of adjacent pucker-pucker interactions allows AC PNTs to accommodate larger stretching. Moreover, the applied tensile strain mostly stretches the $A_I$ and $A_O$ bonds oriented along the direction of the axial strain (see Figure 8(c)). Since at zero strain, the length of the $A_I$-bonds is shorter for AC nanotubes with the smaller diameter (see Figure 4(a)), as a result, a larger tensile strain is needed to reach the critical failure strain, causing the bond fracture.

For ZZ PNTs, the critical strain also decreases with the tube diameter from $\varepsilon_{cr}\approx 0.66$ to $\varepsilon_{cr}\approx 0.55$ (see Figure 9(a), blue squares). In contrast to AC PNTs, tensile strain applied to ZZ PNTs predominantly elongates the B-bonds, which are in parallel with the tube axis, and to a less extent, the $A_I$ and $A_O$ bonds as shown in **Error! Reference source not found.**. Akin to AC PNTs, the bond length of these bonds shortens as the tube diameter decreases (see Figure 4(b)). Therefore, ZZ PNTs with smaller diameters can be elongated considerably longer than ZZ PNTs with larger diameters before their stretched B-bonds reach the maximally allowed bond length limit.

The failure stress as a function of PNT diameter is shown in Figure 9(b). For both AC and ZZ nanotube geometries, the failure stress decreases as the tube diameter increases. The maximal failure stress for AC PNTs is $\sigma_{cr}\approx 21$ GPa, while for ZZ PNTs, it is $\sigma_{cr}\approx 9$ GPa, which is approximately half that of the AC PNT value. The minimal value for the failure stress of AC PNTs is $\sigma_{cr}\approx 13$ GPa, while for ZZ PNTs, it is $\sigma_{cr}\approx 4$ GPa, which is approximately three times lower than that of the AC PNT value.

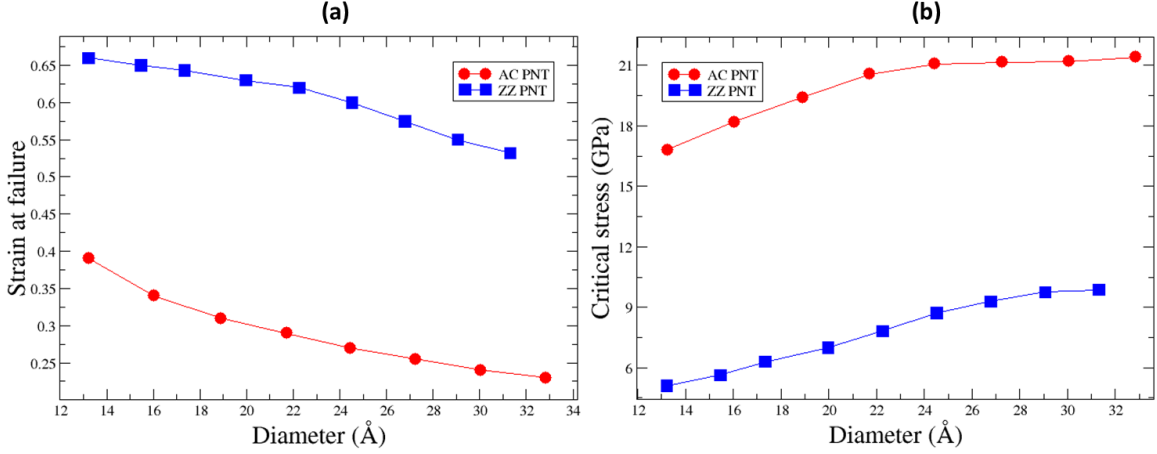

**Figure 9:** (a) Strain at failure (critical strain) plotted against diameter of AC (red circles) and ZZ (blue circles) PNTs; (b) critical (failure) stress plotted vs. diameter of AC (red circles) and ZZ (blue circles) PNTs.

# 4. Conclusions

Applying DFTB method, we examined the elastic properties, deformation and failure of AC and ZZ PNTs subjected to uniaxial tensile strain. We optimized the geometry of the constructed PNTs and studied the curvature effects on their energetics. The obtained energies, which are inversely proportional to the square of tube diameter, are in good agreement with the classical elasticity theory. The Young's modulus, flexural rigidity, radial and thickness Poisson ratios are Y=114.2GPa, δ=0.019 eV·nm$^2$, $\nu_D$ =0.47 and $\nu_t$ =0.11 for AC PNTs, and Y=49.2GPa, δ=0.071 eV·nm$^2$, $\nu_D$ =0.07 and $\nu_t$ =0.21 for ZZ PNTs, respectively. They are relatively insensitive to the tube diameter.

We identified three deformation phases of ZZ PNTs subjected to applied strain: linear elastic (ε ≲ 0.1), bond rotation (0.1 ≲ ε ≲ 0.4) and bond straining phase (ε≳0.4). The linear elastic phase is associated with the interaction of the nearby puckers. In the bond rotation phase the folded configuration of the ZZ PNTs is gradually unfolded. In the bond straining phase the bond length of the B-bonds grows steeply and finally exceeds the maximal limit, resulting in the failure of ZZ PNTs at ε$_{cr}$≈0.6. For AC PNTs, the applied strain leads from the outset to stretching of the A$_I$ and A$_O$ bonds, connecting the P-atoms within the inner and outer tube shells. A substantial fraction of these bonds fractures concurrently at ε$_{cr}$≈0.35, triggering the breakup of AC PNTs.

The failure strain and stress of PNTs are sensitive to the tube diameter. For AC PNTs, the failure strain reduces from ε$_{cr}$≈0.40 to ε$_{cr}$≈0.25, while the failure stress rises from σ$_{cr}$≈13 GPa to σ$_{cr}$≈21 GPa, when the tube diameter increases from D=13.3 Å to D=32.8 Å; for ZZ PNTs, the failure strain diminishes from ε$_{cr}$≈0.66 to ε$_{cr}$≈0.55, while the failure stress increases from σ$_{cr}$≈4 GPa to σ$_{cr}$≈9 GPa, when the tube diameter increases from D=13.2 Å to D=31.1 Å.

In our papers, we studied the mechanical properties of PNTs, which are closely connected to the electronic and thermal properties of PNTs. The special puckered structure of PNTs endows them with high anisotropic electrical and thermal conductivity, which can be tuned by mechanical deformation. Such properties are of great interest for applications of PNTs as building blocks in nanoelectromechanical systems (nanoresonators, nanoaccelerometers and integrated detection devices), electronic circuits (nanotube-based transistors, field emitters and sensors), smart materials and biomedical applications. It stands to reason that in these applications, the knowledge of their mechanical properties, namely elastic constants, deformation and failure mechanism, is indispensable for suitable design and usage of PNTs. The knowledge of the deformation and failure mechanisms sets the limits for mechanical strains and stresses in nanoelectromechanical systems containing PNTs as building blocks, and hence serves as a guideline for designing stable, reliable and robust nano-devices. It is expected that the present findings provide suitable guidelines for the designs and applications of PNTs as building blocks in nano-devices.

## 5. Acknowledgements

The authors want to express their gratitude for the financial funding from A*STAR, Singapore and the usage of supercomputing services at ACRC, Singapore. This work was supported by the A*STAR Computational Resource Centre through the use of its high performance computing facilities.